\documentclass[preprint]{revtex4-1}
\usepackage[english]{babel}
\usepackage{graphicx}

\begin{document}
\title{Random sequential adsorption of trimers and hexamers}
\author{Micha\l{} Cie\'sla$^{1}$}
\email{michal.ciesla@uj.edu.pl}
\author{Jakub Barbasz$^{1,2}$}
\email{ncbarbas@cyf-kr.edu.pl}
\affiliation{
$^1$ M. Smoluchowski Institute of Physics, Jagiellonian University, 30-059 Kraków, Reymonta 4, Poland. \\
$^2$ Institute of Catalysis and Surface Chemistry, Polish Academy of Sciences, 30-239 Kraków, Niezapominajek 8, Poland.
}%
\begin{abstract}
Adsorption of trimers and hexamers built of identical spheres was studied numerically using the Random Sequential Adsorption (RSA) algorithm. Particles were adsorbed on a two dimensional, flat and homogeneous surface. Numerical simulations allow to establish the maximal random coverage ratio, RSA kinetics as well as the Available Surface Function (${\rm ASF}$), which is crucial for determining kinetics of the adsorption process obtained experimentally. Additionally, the density autocorrelation function was measured. All the results were compared with previous results obtained for spheres, dimers and tetramers.
\end{abstract}

\date{\today}

\maketitle

\section{Introduction}
The high interest in irreversible adsorption results from its numerous practical applications in many fields such as medicine and material sciences as well as pharmaceutical and cosmetic industries. Adsorption is crucial for blood coagulation~\cite{bib:Ekdahl2013}, inflammatory response~\cite{bib:Lu2013}, plaque formation~\cite{bib:Gallet2011}, fouling of contact lenses~\cite{bib:Luensmann2012} as well as for ultrafiltration and the operation of membrane filtration units~\cite{bib:Pagana2011}. Controlled adsorption is prerequisite for efficient chromatographic separation and purification, and gel electrophoresis.
\par
The simplest algorithm used for numerical modelling of irreversible adsorption processes is Random Sequential Adsorption (RSA) introduced by Feder \cite{bib:Feder1980}. At first, it was used to model spherical molecules adsorption, but soon it has been used also for modelling more complex particles like ellipsoids, spherocilinders and so on, e.g. \cite{bib:Talbot1989,bib:Vigil1989,bib:Tarjus1991,bib:Viot1992,bib:Ricci1992,bib:Sikorski2011,bib:Pawlowska2013}. Only recently, it has been shown, however, that, for the purposes of adsorption modelling, complex molecules can be successfully approximated using coarse-grain models \cite{bib:Rabe2011,bib:Finch2012,bib:Katira2012,bib:Adamczyk2012rev}. For example, a coarse-grain model of fibrinogen can successfully explain the density of an adsorbed monolayer for a wide range of experimental conditions \cite{bib:Adamczyk2010,bib:Adamczyk2011,bib:Ciesla2013fib}. 
\par
This study focuses on the RSA of trimers and hexamers built of identical spheres on a flat and homogeneous two dimensional surface. There are at least two reasons to justify making a study of this subject. Firstly, trimers and hexamers are the only basic structures, which have not been analysed using coarse-grain models and RSA yet, despite the interest in simpler models like dimers \cite{bib:Ciesla2012dim}, tetramers \cite{bib:Ciesla2013tet} or polymers \cite{bib:Ciesla2013pol}. This work simply completes the library of RSA properties for common basic structures. Secondly, it has been shown that RSA kinetics for tetramers is similar to the one observed for anisotropic molecules and different to that for spheres \cite{bib:Ciesla2013tet}. As a sphere is a better approximation of a hexamer than a tetramer it would be interesting to explore its RSA kinetics. The primary aim of this paper is to find the saturated random coverage ratio of monolayers built as a result of the irreversible trimer and hexamer adsorption. Additionally, we want to determine available surface function, which is crucial for estimating the kinetics of the adsorption process.
\section{Model}
\label{sec:Model}
\begin{figure}[htb]
\centerline{%
\includegraphics[width=6cm]{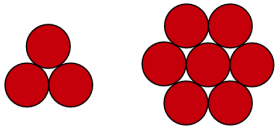}}
\caption{Models of a trimer (left) and a hexamer (right). All spheres (monomers) have radius $r_0$.}
\label{fig:model}
\end{figure}

The model of a trimer and a hexamer consists of three or seven identical spheres, respectively, as shown in Fig.\ref{fig:model}. The radius of a single sphere is $r_0$, and it functions as a length unit. 
\par 
The molecules are placed on a square flat collector surface according to the Random Sequential Adsorption (RSA) algorithm \cite{bib:Feder1980}, described in details elsewhere \cite{bib:Ciesla2013tet}. The algorithm iteratively attempts to put randomly oriented and positioned particle on the collector. If this particle does not overlap with previously adsorbed particles it is irreversibly adsorbed and holds its place till the end of the simulation. If there is an overlap, the particle is removed and abandoned. The number of algorithm iterations $N$ is commonly expressed using dimensionless time:
\begin{equation}
\label{eq:dimlesstime}
t = N\frac{S_{\rm M}}{S_{\rm C}},
\end{equation}
where $S_{\rm M}$ is an area covered by a single trimer ($3\pi r_0^2$) or hexamer ($7\pi r_0^2$) and $S_{\rm C}$ is a collector size. In case of these simulations, square collectors of  side $1000r_0$ were used, so $S_{\rm C} = 10^6 r_0^2$. Simulations were run until $t = 10^5$, which corresponds to $N = 10^5\frac{10^6 r_0^2}{S_{\rm M}}$ algorithm steps.
\par
During simulation, the current coverage ratio $\theta(t)$ was monitored:
\begin{equation}
\theta(t) = n(t) \frac{S_{\rm M}}{S_{\rm C}},
\end{equation}
where $n(t)$ is a number of adsorbed particles after the number of steps corresponding to dimensionless time $t$. To decrease statistical error, $100$ independent RSA simulations were performed for each model.
\section{Results and discussion}
\begin{figure}[htb]
\centerline{%
\includegraphics[width=6cm]{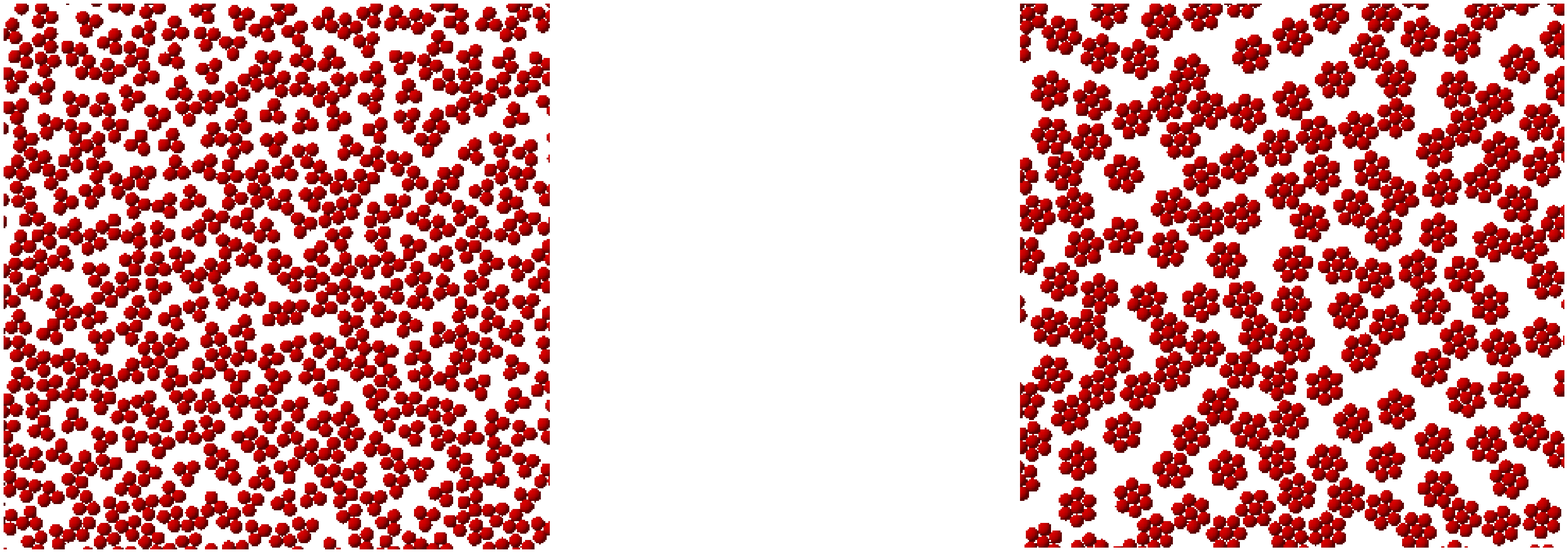}}
\caption{Example coverages built of trimers (left) and hexamers (right).}
\label{fig:examples}
\end{figure}

Obtained example coverages are presented in Fig.\ref{fig:examples}.
Saturated coverage ratio, which is one of the most important characteristic of an adsorption layer, is reached, in general, after an infinite number of RSA iterations: $\theta_{\rm max} \equiv \theta(t\to\infty)$. Therefore, to determine it from a finite time simulation the RSA kinetics model has to be used.  
\subsection{Kinetics of the Random Sequential Adsorption}
For spheres the kinetics of the RSA obeys the Feder law \cite{bib:Swendsen1981,bib:Privman1991}:
\begin{equation}
\theta_{\rm max} - \theta(t) \sim t^{-1/d},
\label{fl}
\end{equation}
where $d$ is a collector dimension and $t$ is a dimensionless time (\ref{eq:dimlesstime}). Relation (\ref{fl}) has been tested numerically and appears to be valid for one to six dimensional collectors~\cite{bib:Torquato2006} as well as for fractal collectors having $0<d<3$~\cite{bib:Ciesla2012fractal,bib:Ciesla2013sponge}. For different adsorbates:  ellipsoids, dimers and polymers, it is also valid; however, parameter $d$ depends also on particle anisotropy and their number of degrees of freedom \cite{bib:Ciesla2013pol,bib:Hinrichsen1986}. For example for dimers, tetramers and stiff elongated particles adsorption on two dimensional surface, $d \approx 3$.
\par
\begin{figure}[htb]
\vspace{1cm}
\centerline{%
\includegraphics[width=6cm]{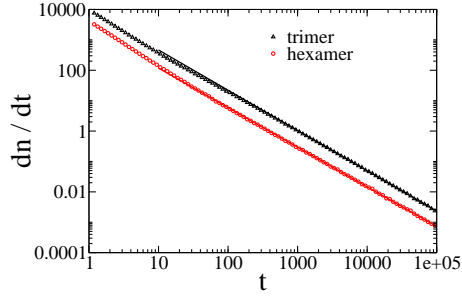}}
\caption{Dependence of time derivative of the mean number of adsorbed particles on the dimensionless time. Triangles and circles are simulation data for trimer and hexamer, respectively, whereas solid lines correspond to power fits obtained for $t>1000$: $dn / dt = 9865.7 t^{-1.325}$ for the trimer model and  $dn / dt = 2375.6  t^{-1.301}$ for the hexamer model. Determined values of exponent $d$ in Eq.(\ref{fl}) are as follows: $d=3.08$ and $d=3.32$ for trimer and hexamer model, respectively.}
\label{fig:dnt}
\end{figure}

For large enough time $t$, the exponent in Eq.\ref{fl} can be measured directly from $d\theta / dt$ dependence on $t$ using the least squares approximation method (see Fig.\ref{fig:dnt}). 
For both, trimer and hexamer, the obtained values of parameter $d$ are significantly larger than $2$, the value expected for spherical particles. This is particularly surprising in case of hexamer adsorption, for which the shape anisotropy is very small, and $d$ is bigger than for a trimer.
\subsection{Saturated random coverage ratio}
Having determined the RSA kinetics, Eq.(\ref{fl}) can be rewritten as $\theta(y) = \theta_{\rm max} - A y$, where $A$ is a constant coefficient and $y=t^{-1/d}$. Saturated random coverage ratio $\theta_{\rm max}$ is obtained by a linear approximation of this relation for $y=0$. Here, $\theta_{\rm max}=0.5234$ and $\theta_{\rm max}=0.4920$ for trimers and hexamers, respectively. The relative error for both the values is approximately $0.5\%$. It originates mainly from the statistics: the standard deviation of deposited particles at the end of a simulation, as well as from error of adsorption kinetics fit. Obtained ratio is smaller than $\theta_{\rm max} \approx 0.54$ obtained for spheres \cite{bib:Torquato2006}, dimers \cite{bib:Ciesla2012dim} or very short polymers \cite{bib:Ciesla2013pol}. The $\theta_{\rm max}$ for trimers is, however, very similar to the value obtained for the rhomboid model of a tetramer \cite{bib:Ciesla2013tet}.
\subsection{Adsorption kinetics}
\begin{figure}[htb]
\vspace{1cm}
\centerline{%
\includegraphics[width=6cm]{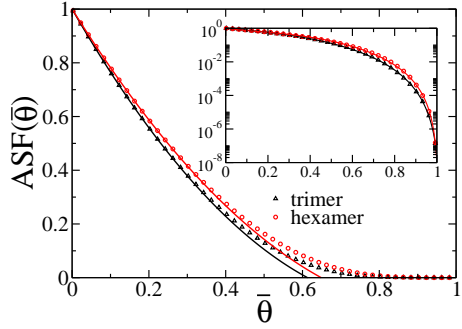}}
\caption{Dependence of Available Surface Function on a normalised coverage ratio. Triangles and circles are simulation data for the trimer and hexamer model, respectively. Solid lines correspond to the fits (\ref{eq:asffit}) obtained for $\bar{\theta}<0.2$: ${\rm ASF}(\theta) = 1 - 4.755 \, \theta + 5.107 \, \theta^2$ for the trimer adsorption and  ${\rm ASF}(\theta) = 1 - 4.664 \, \theta + 4.797 \, \theta^2$ for the hexamer adsorption. Inset shows the same data in logarithmic scale, and fits there correspond to: 
${\rm ASF}(\bar{\theta}) = \left( 1 - 1.315 \, \bar{\theta} + 4.007 \, \bar{\theta}^2 - 5.977 \, \bar{\theta}^3 \right) \left(1-\bar{\theta}^4 \right)$ for trimer and ${\rm ASF}(\bar{\theta}) = \left( 1 - 0.121 \, \bar{\theta} + 10.884 \, \bar{\theta}^2 - 8.684 \, \bar{\theta}^3 \right) \left(1-\bar{\theta}^4 \right)$ for hexamer model.
}
\label{fig:asf}
\end{figure}

Kinetics of the adsorption process depends on two factors. The first one is transport process, which shifts particles close to the surface or interface where they are adsorbed. As it depends on a specific experimental conditions, it hardly becomes a subject of the general theoretical analysis. The second factor is probability of adsorption, which changes with diminishing area of uncovered surface. The dependence between adsorption probability and temporary coverage ratio is defined as Available Surface Function (${\rm ASF}$) and it can be easily determined from the RSA simulation. Fig. \ref{fig:asf}. shows the ${\rm ASF}$  dependence on normalised coverage $\bar{\theta} = \theta / \theta_{\rm max}$.
For almost empty collector (small $\theta$), the probability decreases linearly because each successful adsorption act blocks a specified amount of the collector surface. When $\theta$ grows, these blocked areas start to overlap, which slows down the rate of probability decrease. Therefore for low values of $\theta$, the ${\rm ASF}(\theta)$ is typically approximated by a quadratic fit \cite{bib:Tarjus1991,bib:Ricci1992,bib:AdamczykBook}:
\begin{equation}
{\rm ASF}(\theta) = 1 - C_1 \theta + C_2 \theta^2 + o(\theta^2),
\label{eq:asffit}
\end{equation}
where the expansion coefficient $C_1$ corresponds to the surface area blocked by a single particle, whereas $C_2$ denotes a cross-section of the area blocked by two independent molecules. Both of them are directly related to the second $B_2=1/2C_1$ and third $B_3=1/3C_1^2-2/3C_2$ viral coefficients of the equilibrium trimer or hexamer monolayer \cite{bib:Tarjus1991,bib:AdamczykBook}. For example, the 2D pressure $P$ and the chemical potential of particle $\mu$ can be expressed via the series expansion at a low coverage limit \cite{bib:AdamczykBook}
\begin{equation}
\begin{array}{c}
P = \frac{k_{\rm B} T}{S_{\rm F}} \left( \theta + B_2\theta^2 + B_3\theta^3 + o(\theta^3) \right), \\
\mu = \mu_0 + k_{\rm B} T \left( \ln \theta  + 2B_2\theta + \frac{3}{2}B_3 \theta^3 + o(\theta^3)   \right), 
\end{array}
\end{equation}
where $k_B$ is the Boltzmann constant, $T$ is the absolute temperature,
and $\mu_0$ is the reference potential.
\par
Results presented in Fig.\ref{fig:asf} show that $C_1$ for both particle types is approximately $15\%$ bigger than for a spherical particle, for which $C_1=4$. Parameter $C_2$ is almost $50\%$ larger than for spheres ($C_2 \approx 3.308$), which, probably, is the result of more irregular shape as this parameter is significantly bigger for the trimer model than for the hexamer model.
\par
The saturation limit of the ${\rm ASF}$, is more important for adsorption kinetics calculations \cite{bib:Adamczyk2010,bib:Ciesla2013dim}, is, for particles characterised by exponent $d \approx 3$, typically approximated by \cite{bib:Ricci1992}:
\begin{equation}
\label{asfjfit}
{\rm ASF}(\bar{\theta}) = (1 + a_1 \bar{\theta} + a_2 \bar{\theta}^2 + a_3 \bar{\theta}^3)(1-\bar{\theta})^4.
\end{equation}
The  last factor in the above equation is directly related to Eq.\ref{fl} as the adsorption probability is proportional to the growth rate of $\theta$. Exponent $4$ here results from $d \approx 3$. As shown in the Fig.\ref{fig:asf} inset, the above relation is also valid for trimer and hexamer adsorption.  
\subsection{Coverage structure}
Although the saturated random coverage ratio is the main characteristic of RSA monolayer structure it only carries the information about mean density of adsorbed particles. To get deeper insight into the structure of RSA monolayers, we measured density fluctuations as well as density autocorrelations for obtained coverages.
\subsection{Density fluctuations}
\begin{figure}[htb]
\vspace{1cm}
\centerline{%
\includegraphics[width=6cm]{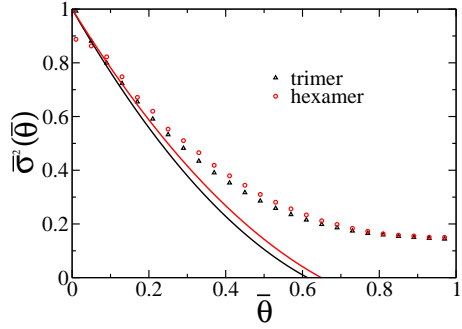}}
\caption{Dependence of normalised density variance on coverage ratio. Triangles and circles are measured values for the trimer and the hexamer model, respectively; whereas solid lines correspond to the ${\rm ASF}$ fit in a low coverage limit (\ref{eq:asffit}).}
\label{fig:varasf}
\end{figure}

The typical experimental procedure used for estimation of density fluctuations of adsorbed particles, described by Adamczyk et al. \cite{bib:Adamczyk1996}, can be used also for monolayers generated by the RSA. Here, for a given coverage ratio $\theta$, the collector has been divided into square boxes containing at average $10$ particles. The normalized variance of the number of particles in the box $n_B$ for the given coverage $\theta$: $\bar{\sigma}^2(\theta) = \sigma^2\left( n_B(\theta) \right) / \langle n_B(\theta) \rangle$ is used as a density fluctuation measure. Its dependence on the coverage ratio $\theta$ is shown in Fig.\ref{fig:varasf}.
It is worth to notice that in a limit of small coverages ${\rm ASF}(\theta) = \bar{\sigma}^2(\theta)$ \cite{bib:Schaaf1995} . Plots in Fig.\ref{fig:varasf} confirm this agreement for $\bar{\theta}<0.2$.
\subsection{Density autocorrelation}
\begin{figure}[htb]
\vspace{1cm}
\centerline{%
\includegraphics[width=6cm]{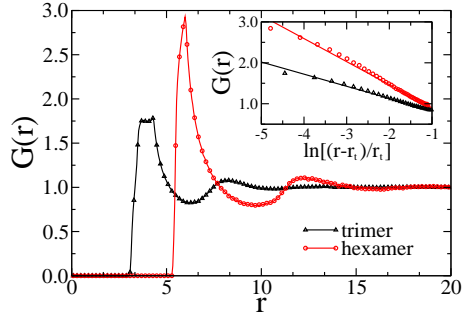}}
\caption{Density autocorrelation function G(r) for trimer and hexamer models. Inset shows a logarithmic singularity at $r \to r_t+$, where $r_t$ is a maximum of $G(r)$. Parameter $r_t=4.28$ and $r_t=5.98$ for the trimer and hexamer model, respectively.}
\label{fig:cor}
\end{figure}

The density autocorrelation function is defined as:
\begin{equation}
G(r) = \frac{P(r)}{2\pi r \rho},
\end{equation}
where $P(r)dr$ is a probability of finding two particles in a distance between $r$ and $r+dr$. Here, the distance $r$ is measured between the geometric centres of molecules.  As $\rho$ is the mean density of particles inside a covering layer, then $G(r\to \infty) = 1$. In case of spherical particles, $G(r)$ has a logarithmic singularity in the touching limit \cite{bib:Swendsen1981} and superexponential decay at large distances \cite{bib:Bonnier1994}. Density autocorrelation functions for trimer and hexamer monolayers are shown in Fig.\ref{fig:cor}.
Due to different size of particles, the density autocorrelation function for hexamers is shifted right compared to the one for trimers. For trimers the first maximum is wide due to particle shape anisotropy. Therefore, for dense packing, the distance between closest particles varies because it depends on trimers' relative orientations. For hexamers of significantly smaller shape anisotropy, the density autocorrelation function looks similar to the one for spherical particles and the analytically predicted logarithmic singularity in the touching limit \cite{bib:Swendsen1981} can be clearly seen. It can also be observed for trimers, when restricting to the region to the right off the flat maximum. For large $r$, autocorrelations approach the mean density value very fast, similarly as for spheres and other particles like dimers or tetramers.
\section{Summary}
The saturated random coverage ratio of a trimer monolayer is $\theta_{\rm max}=0.5234$. It is smaller than the one for spheres and dimers and similar to the obtained for rhomboid tetramers \cite{bib:Ciesla2013tet}. The saturated coverage ratio for hexamers is $\theta_{\rm max}=0.4920$ and it is similar to the value obtained for the square model of a tetramer \cite{bib:Ciesla2013tet}. At jamming limit, the kinetics of RSA of trimers and hexamers shows behaviour typical of anisotropic molecules, which is highly unexpected especially for hexamers, considering their small shape anisotropy. Properties of the density autocorrelation function in dense monolayers are, in general, similar to the observed for spheres monolayers. 
\par 
This work completes the analysis of RSA monolayers built of basic particles composed of identical spheres. Therefore, for convenience, Table \ref{tab:summary} presents together the most important parameters of such monolayers, based on the results of this work and of \cite{bib:Ciesla2012dim,bib:Ciesla2013tet,bib:Ciesla2013pol,bib:Torquato2006}.
\begin{table}
\centerline{
\begin{tabular}{|c|c|c|c|c|}
\hline
particle type & $\theta_{\rm max}$ & $d$ & $C_1$ & $C_2$ \\
\hline
\hline
sphere \cite{bib:Feder1980,bib:Ciesla2013pol} & $0.545$ & $2.0$ & $4.0$ & $\frac{6\sqrt{3}}{\pi} \approx 3.308$ \\
dimer \cite{bib:Ciesla2012dim,bib:Ciesla2013pol} & $0.541$ & $2.8$ & $4.84$ & $5.49$ \\
3-chain \cite{bib:Ciesla2013pol} & $0.542$ & $4.2$ & $5.28$ & $6.56$ \\
trimer & $0.523$ & $3.1$ & $4.76$ & $5.11$ \\
4-chain \cite{bib:Ciesla2013pol} & $0.543$ & $6.0$ & $5.54$ & $7.31$ \\
tetramer (rhomboid) \cite{bib:Ciesla2013tet} & $0.521$ & $3.4$ & $4.74$ & $5.09$ \\
tetramer (square) \cite{bib:Ciesla2013tet} & $0.491$ & $3.3$ & $4.84$ & $5.22$ \\
hexamer & $0.492$ & $3.3$ & $4.66$ & $4.80$ \\
6-chain \cite{bib:Ciesla2013pol} & $0.548$ & $9.8$ & $5.78$ & $8.18$ \\
\hline
\end{tabular}
}
\caption{Saturated random coverage ratio $\theta_{\rm max}$, RSA kinetics exponent $d$ and ${\rm ASF}$ low coverage limit coefficient for the most common models of particles.} 
\label{tab:summary}
\end{table}
\section*{Acknowledgement}
This work was supported by Polish National Science Center \newline grant no.  UMO-2012/07/B/ST4/00559.
%

\end{document}